\documentclass[aps,superscriptaddress,nofootinbib,floatfix,notitlepage]{revtex4-1}
\usepackage[utf8x]{inputenc}
\pdfoutput=1
\usepackage{graphicx}
\usepackage{hyperref}
\usepackage{bm}

\begin{document}

\title{The \textbf{UT}\textit{fit} Collaboration Average of \boldmath$D$ meson
  mixing data: Winter 2014
  \vspace*{0.5cm}
}

\collaboration{\begin{figure}[h!]
  \begin{center}
  \includegraphics[width=0.13\textwidth]{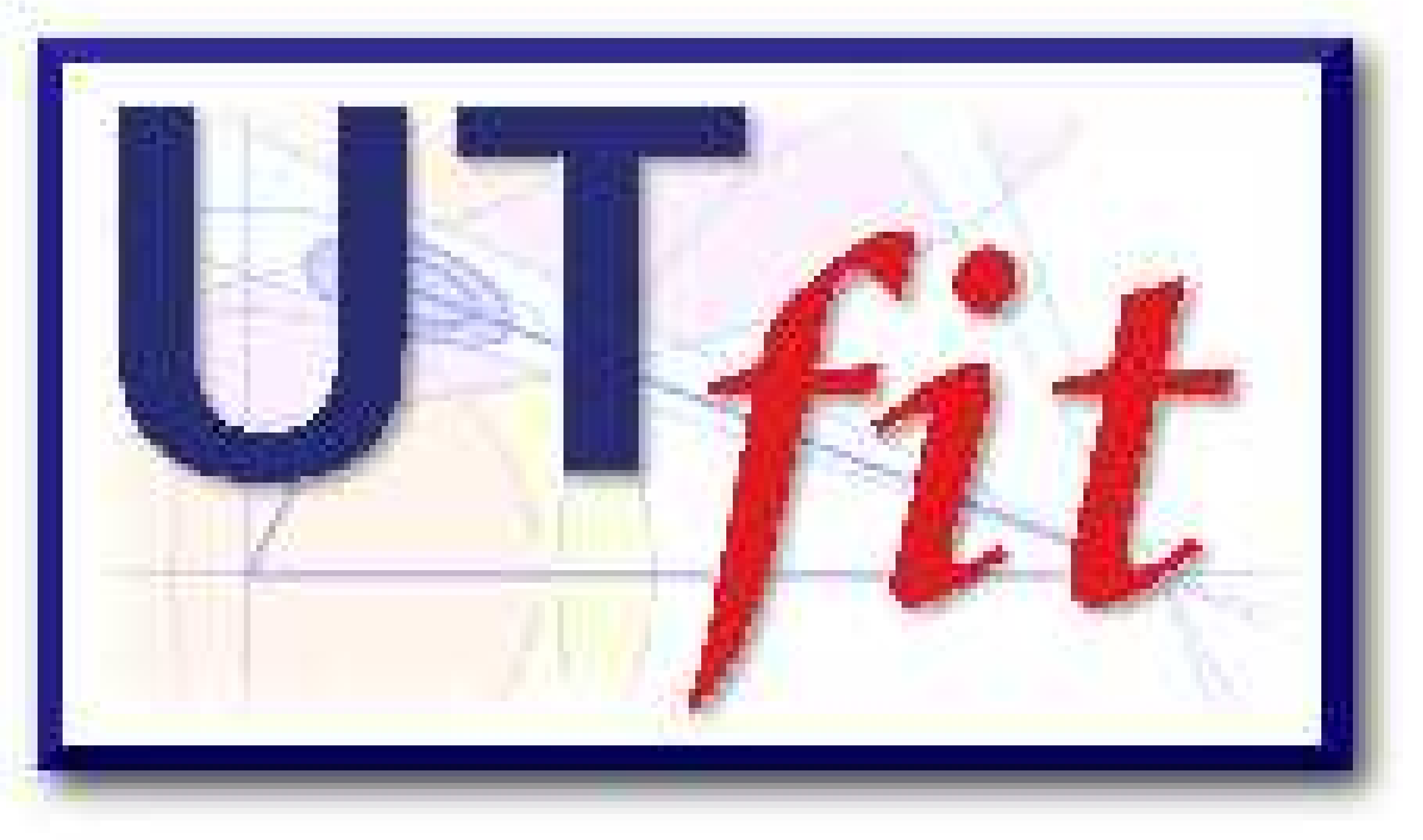}
  \end{center}
 \end{figure}\textbf{UT}\textit{fit} Collaboration
}
\homepage{http://www.utfit.org} 
\author{A.J.~Bevan}
\affiliation{Queen Mary, University of London, Mile End Road, London E1 4NS, United Kingdom}
\author{M.~Bona}
\affiliation{Queen Mary, University of London, Mile End Road, London E1 4NS, United Kingdom}
\author{M.~Ciuchini}
\affiliation{INFN,  Sezione di Roma Tre, Via della Vasca Navale 84, I-00146 Roma, Italy}
\author{D.~Derkach}
\affiliation{Department of Physics, University of Oxford, OX1 3PN Oxford, United Kingdom}
\author{E.~Franco}
\affiliation{INFN, Sezione di Roma, Piazzale A. Moro 2, I-00185 Roma, Italy}
\author{V.~Lubicz}
\affiliation{INFN,  Sezione di Roma Tre, Via della Vasca Navale 84, I-00146 Roma, Italy}
\affiliation{Dipartimento di Matematica e Fisica, Universit{\`a} di Roma Tre, Via della Vasca Navale 84, 
  I-00146 Roma, Italy} 
\author{G.~Martinelli}
\affiliation{INFN, Sezione di Roma, Piazzale A. Moro 2, I-00185 Roma, Italy}
\affiliation{SISSA-ISAS, Via Bonomea 265, I-34136 Trieste, Italy} 
\author{F.~Parodi}
\affiliation{Dipartimento di Fisica, Universit\`a di Genova and INFN, Via Dodecaneso 33, I-16146
  Genova, Italy} 
\author{M.~Pierini}
\affiliation{CERN, CH-1211 Geneva 23, Switzerland}
\author{C.~Schiavi}
\affiliation{Dipartimento di Fisica, Universit\`a di Genova and INFN, Via Dodecaneso 33, I-16146
  Genova, Italy} 
\author{L.~Silvestrini}
\affiliation{INFN, Sezione di Roma, Piazzale A. Moro 2, I-00185 Roma, Italy}
\author{V.~Sordini}
\affiliation{IPNL-IN2P3, 4 Rue Enrico Fermi, F-69622 Villeurbanne Cedex, France}
\author{A.~Stocchi}
\affiliation{Laboratoire de l'Acc\'el\'erateur Lin\'eaire, IN2P3-CNRS et
  Universit\'e de Paris-Sud, BP 34, F-91898 Orsay Cedex, France}
\author{C.~Tarantino}
\affiliation{INFN,  Sezione di Roma Tre, Via della Vasca Navale 84, I-00146 Roma, Italy}
\affiliation{Dipartimento di Matematica e Fisica, Universit{\`a} di Roma Tre, Via della Vasca Navale 84,
  I-00146 Roma, Italy} 
\author{V.~Vagnoni}
\affiliation{INFN, Sezione di Bologna,  Via Irnerio 46, I-40126 Bologna, Italy}

\begin{abstract}
\phantom{a}
\vspace{0.5cm}
  We update the analysis of $D$ meson mixing including the latest
  experimental results as of January 2014. We derive constraints on
  the parameters $M_{12}$, $\Gamma_{12}$ and $\Phi_{12}$ that describe
  $D$ meson mixing using all available data, allowing for CP
  violation. We also provide posterior distributions
  for observable parameters appearing in $D$ physics.
\end{abstract}
 
\maketitle

Almost two years ago, we presented our combination of the $D$ mixing
experimental data, yielding a quite precise determination of the
mixing parameters showing no sign of CP
violation~\cite{Bevan:2012waa}.  Recently, the LHCb Collaboration has
improved several important
measurements~\cite{Aaij:2013ria,Aaij:2013wda}, and updates have also
come from the other
experiments~\cite{Lees:2012qh,Staric:2012ta,Asner:2012xb,Aaltonen:2013pja}.
These improvements result in a remarkable accuracy in the
determination of the CP violating phase in charm mixing, implying
strong contraints on possible extensions of the Standard Model (SM).
An update of our fit is timely and can be of use for phenomenological
analyses of physics beyond the SM.

\begin{table}[hp]
  \centering
  \begin{tabular}{|cccccccc|}
    \hline
    Observable & Value & \multicolumn{5}{c}{Correlation Coeff.} &
    Reference\\ \hline
    $y_{CP}$ & $(0.866 \pm 0.155)\%$ & & & & & &
    \cite{Link:2000cu,Csorna:2001ww,Zupanc:2009sy,Aaij:2011ad,Lees:2012qh,Staric:2012ta}
    \\ \hline
    $A_\Gamma$ & $(-0.014 \pm 0.052)\%$ & & & & & &
    \cite{Aitala:1999dt,Lees:2012qh,Staric:2012ta,Aaij:2013ria}
    \\ \hline
    $x$ & $(0.79 \pm 0.29 \pm 0.08 \pm 0.12)\%$ & 1 & -0.007 & -0.255$\alpha$ &
    0.216 & & \cite{Abe:2007rd}\\
    $y$ & $(0.30 \pm 0.24 \pm 0.1 \pm 0.07)\%$ & -0.007 & 1 & -0.019$\alpha$ &
    -0.280 & & \cite{Abe:2007rd}\\
    $\vert q/p \vert$ & $(0.96 \pm 0.21)$ & -0.255$\alpha$
    & -0.019$\alpha$ & 1 &
    -0.128$\alpha$ & & \cite{Abe:2007rd}\\
    $\phi$ & $(-2.5 \pm 10.5)^\circ$ & 0.216
    & -0.280 &
    -0.128$\alpha$ & 1 & & \cite{Abe:2007rd}\\\hline
    $x$ & $(0.16 \pm 0.23 \pm 0.12 \pm 0.08)\%$ & 1 & 0.0615 & & & &
    \cite{delAmoSanchez:2010xz}\\
    $y$ & $(0.57 \pm 0.20 \pm 0.13 \pm 0.07)\%$ & 0.0615 & 1 & & & &
    \cite{delAmoSanchez:2010xz}\\
    \hline
    $R_M$ & $(0.0130 \pm 0.0269)\%$ & & & & & &
    \cite{Aitala:1996vz,Cawlfield:2005ze,Aubert:2004bn,Aubert:2007aa,Bitenc:2008bk}
    \\\hline
    $(x^\prime_+)_{K\pi\pi}$ & $(2.48 \pm 0.59 \pm 0.39)\%$ & 1 & -0.69 &  &   &
    & \cite{Aubert:2008zh}\\
    $(y^\prime_+)_{K\pi\pi}$ & $(-0.07 \pm 0.65 \pm 0.50)\%$ & -0.69 &
    1 &  &   &
    & \cite{Aubert:2008zh}\\
    $(x^\prime_-)_{K\pi\pi}$ & $(3.50 \pm 0.78 \pm 0.65)\%$ & 1 & -0.66 &  &   &
    & \cite{Aubert:2008zh}\\
    $(y^\prime_-)_{K\pi\pi}$ & $(-0.82 \pm 0.68 \pm 0.41)\%$ & -0.66 &
    1 &  &   &
    & \cite{Aubert:2008zh}\\ \hline
    $R_D$ & $(0.533 \pm 0.107 \pm 0.045)\%$ & 1 & 0 & 0 & -0.42 & 0.01
    & \cite{Asner:2012xb}\\
    $x^2$ & $(0.06 \pm 0.23 \pm 0.11)\%$ & 0 & 1 & -0.73 & 0.39
    & 0.02
    & \cite{Asner:2012xb}\\
    $y$ & $(4.2 \pm 2 \pm 1)\%$ & 0. & -0.73 & 1 & -0.53  & -0.03
    & \cite{Asner:2012xb}\\ 
    $\cos\delta_{K\pi}$ & $(0.84 \pm 0.2 \pm 0.06)$ & -0.42 &
    0.39 & -0.53 & 1  & 0.04
    & \cite{Asner:2012xb}\\
    $\sin\delta_{K\pi}$ & $(-0.01 \pm 0.41 \pm 0.04)$ & 0.01
    & 0.02 & -0.03 & 0.04  & 1
    & \cite{Asner:2012xb}\\
   \hline
    $R_D$ & $(0.3030 \pm 0.0189)\%$ & 1 & 0.77 & -0.87 &   &
    & \cite{Aubert:2007wf}\\
    $(x^\prime_+)^2_{K\pi}$ & $(-0.024 \pm 0.052)\%$ & 0.77 & 1 & -0.94 &   &
    & \cite{Aubert:2007wf}\\
    $(y^\prime_+)_{K\pi}$ & $(0.98 \pm 0.78)\%$ & -0.87 & -0.94 & 1  & &
    & \cite{Aubert:2007wf}\\\hline
    $A_D$ & $(-2.1 \pm 5.4)\%$ & 1 & 0.77 & -0.87 &   &
    & \cite{Aubert:2007wf}\\
    $(x^\prime_-)^2_{K\pi}$ & $(-0.020 \pm 0.050)\%$ & 0.77 & 1 & -0.94 &   &
    & \cite{Aubert:2007wf}\\
    $(y^\prime_-)_{K\pi}$ & $(0.96 \pm 0.75)\%$ & -0.87 & -0.94 & 1  & &
    & \cite{Aubert:2007wf}\\\hline
    $R_D$ & $(0.364 \pm 0.018)\%$ & 1 & 0.655 & -0.834 &   &
    & \cite{Zhang:2006dp}\\
    $(x^\prime_+)^2_{K\pi}$ & $(0.032 \pm 0.037)\%$ & 0.655 & 1 & -0.909 &   &
    & \cite{Zhang:2006dp}\\
    $(y^\prime_+)_{K\pi}$ & $(-0.12 \pm 0.58)\%$ & -0.834 & -0.909 & 1  & &
    & \cite{Zhang:2006dp}\\\hline
    $A_D$ & $(2.3 \pm 4.7)\%$ & 1 & 0.655 & -0.834 &   &
    & \cite{Zhang:2006dp}\\
    $(x^\prime_-)^2_{K\pi}$ & $(0.006 \pm 0.034)\%$ & 0.655 & 1 & -0.909 &   &
    & \cite{Zhang:2006dp}\\
    $(y^\prime_-)_{K\pi}$ & $(0.20 \pm 0.54)\%$ & -0.834 & -0.909 & 1  & &
    & \cite{Zhang:2006dp}\\\hline
    $R_D$ & $(0.351 \pm 0.035)\%$ & 1 & -0.967 & 0.900 &   &
    & \cite{Aaltonen:2013pja}\\
    $(y^\prime_\mathrm{CPA})_{K\pi}$ & $(0.43 \pm 0.43)\%$ & -0.967 & 1 & -0.975 &   &
    & \cite{Aaltonen:2013pja}\\
    $(x^\prime_\mathrm{CPA})^2_{K\pi}$ & $(0.008 \pm 0.018)\%$ & 0.900 & -0.975 & 1  & &
    & \cite{Aaltonen:2013pja}\\\hline
    $R_D$ & $(0.3568 \pm 0.0058 \pm 0.0033)\%$ & 1 & -0.894 & 0.77 & -0.895  &
    0.772 & \cite{Aaij:2013wda}\\
    $(y^\prime_+)_{K\pi}$ & $(0.48 \pm 0.09 \pm 0.06)\%$ & -0.894 & 1
    & -0.949  & 0.765 & -0.662
    & \cite{Aaij:2013wda}\\
    $(x^\prime_+)^2_{K\pi}$ & $(6.4 \pm 4.7 \pm 3)10^{-5}$ & 0.77 &
    -0.949 & 1 & -0.662  & 0.574
    & \cite{Aaij:2013wda}\\
    $(y^\prime_-)_{K\pi}$ & $(0.48 \pm 0.09 \pm 0.06)\%$ & -0.895 &
    0.765 & -0.662 & 1  & -0.95
    & \cite{Aaij:2013wda}\\
    $(x^\prime_-)^2_{K\pi}$ & $(4.6 \pm 4.6 \pm 3)10^{-5}$ & 0.772 &
    -0.662 & 0.574  & -0.95 & 1
    & \cite{Aaij:2013wda}\\\hline
  \end{tabular}
  \caption{Experimental data used in the analysis, from
    ref.~\cite{[{}][{ and online updates at
        \url{http://www.slac.stanford.edu/xorg/hfag/}}]Amhis:2012bh}. $\alpha
    = (1 + \vert q/p \vert)^2/2$. Asymmetric errors have been 
    symmetrized.}
  \label{tab:dmixexp}
\end{table}

In this letter, we perform a fit to the experimental data in
Table~\ref{tab:dmixexp} following the statistical method described in
ref.~\cite{Ciuchini:2000de} improved with a Markov-chain Monte Carlo
as implemented in the BAT library~\cite{Caldwell:2008fw}. The following
parameters are varied with flat priors in a sufficiently large range:
\begin{equation}
x=\frac{\Delta m}{\Gamma}\,,\qquad y=\frac{\Delta\Gamma}{2\Gamma}\,, \qquad
\left\vert \frac{q}{p}\right\vert\,,\qquad\delta_{K\pi}\,,\qquad \delta_{K\pi\pi}\,,\qquad R_D\,,
\end{equation}
where $q$ and $p$ are defined as $\vert D_{L,S} \rangle = p \vert D^0
\rangle \pm q \vert \bar D^0 \rangle$ with $\vert p \vert^2+\vert q
\vert^2 = 1$, $\delta_{K\pi(\pi)}$ is the strong phase difference
between the amplitudes $A(\bar D\to K^+\pi^-(\pi^0))$ and $A(D\to
K^+\pi^-(\pi^0))$ and
\begin{equation}
  \label{eq:xyandco}
   R_D = \frac{\Gamma(D^0 \to K^+\pi^-)+\Gamma(\bar D^0 \to
     K^-\pi^+)}{\Gamma(D^0 \to K^-\pi^+)+\Gamma(\bar D^0 \to
     K^+\pi^-)}\,.
\end{equation}

We make the following assumptions in order to combine the measurements
in Table~\ref{tab:dmixexp}: i) we assume that Cabibbo
allowed (CA) and doubly Cabibbo suppressed (DCS) decays are purely
tree-level SM processes, neglecting direct CP violation; ii)
we neglect the weak phase difference between these channels, which is
of $\mathcal{O}(10^{-3})$.  One can then write the following
equations~\cite{Branco:1999fs,Raz:2002ms,Ciuchini:2007cw,
  Kagan:2009gb,Grossman:2009mn,Bevan:2012waa}:
\begin{eqnarray}
  \label{eq:xyandco_gen}
  &&\delta = \frac{1 - \vert q/p \vert^2}{1+\vert q/p
    \vert^2} \,,\quad \arg(\Gamma_{12}\, q/p) = \arg (y+i \delta x)\,,\quad
  A_M = \frac{\vert q/p \vert^4 -1}{\vert q/p
    \vert^4+1}\,, \quad
  R_M =\frac{x^2+y^2}{2}\,, \\*
  &&
  \left(
    \begin{array}{c}
      x^\prime_f \\
      y^\prime_f
    \end{array}
  \right) =
  \left(
    \begin{array}{cc}
      \cos \delta_{f} & \sin \delta_{f} \\
      -\sin \delta_{f} & \cos \delta_{f}
    \end{array}
  \right)   \left(
    \begin{array}{c}
      x \\
      y
    \end{array}
  \right)
  \,,\quad
  (x^{\prime}_\pm)_f = \left\vert
    \frac{q}{p}
  \right\vert^{\pm 1}(x^\prime_f\cos \phi \pm y^\prime_f \sin
  \phi)\,,\quad
  (y^\prime_\pm)_f   =
  \left\vert
    \frac{q}{p}
  \right\vert^{\pm 1}(y^\prime_f\cos \phi \mp x^\prime_f \sin
  \phi)\,,\nonumber \\*
  && y_\mathrm{CP} =
  \left(
    \left\vert
      \frac{q}{p}
    \right\vert + \left\vert
      \frac{p}{q}
    \right\vert
  \right) \frac{y}{2} \cos \phi- \left(
    \left\vert
      \frac{q}{p}
    \right\vert - \left\vert
      \frac{p}{q}
    \right\vert
  \right) \frac{x}{2}\sin \phi\,,\quad A_\Gamma =  \left(
    \left\vert
      \frac{q}{p}
    \right\vert - \left\vert
      \frac{p}{q}
    \right\vert
  \right) \frac{y}{2} \cos \phi- \left(
    \left\vert
      \frac{q}{p}
    \right\vert + \left\vert
      \frac{p}{q}
    \right\vert
  \right) \frac{x}{2}\sin \phi\,, \nonumber \\*
  && \left(y^\prime_\mathrm{CPA}\right)_f=\frac{(y^\prime_+)_f+(y^\prime_-)_f}{2}\,,\quad
  \left(x^\prime_\mathrm{CPA}\right)^2_f+\left(y^\prime_\mathrm{CPA}\right)^2_f=\frac{(x^\prime_+)^2_f+(x^\prime_-)^2_f+
    (y^\prime_+)^2_f+(y^\prime_-)^2_f}{2}\,,
  \nonumber
\end{eqnarray}
valid for Cabibbo allowed and doubly Cabibbo suppressed final states. 

In the standard CKM phase convention (taking $\mathrm{CP}\vert D\rangle =
\vert \bar D \rangle$), within the approximation we are using, CA and DCS decay
amplitudes have vanishing weak phase and $\phi = \mathrm{arg}(q/p)$.
Given the present experimental accuracy, one can assume $\Gamma_{12}$ to be
real,\footnote{See ref.~\cite{kagantaucharm} for a discussion of the size
of arg($\Gamma_{12}$).}
leading to the relation 
\begin{equation}
\phi = \mathrm{arg}(y+i \delta x)\,.\label{eq:phi}
\end{equation}

For the purpose of constraining NP, it is useful
to express the fit results in terms of the $\Delta C=2$ effective
Hamiltonian matrix elements $M_{12}$ and $\Gamma_{12}$:
\begin{eqnarray}
  &&\vert M_{12} \vert = \frac{1}{\tau_D } \sqrt{\frac{x^2+\delta^2
      y^2}{4(1-\delta^2)}}\sim \frac{x}{2\tau_D }+\mathcal{O}(\delta^2)\,,\quad
  \vert \Gamma_{12} \vert= \frac{1}{\tau_D }\sqrt{\frac{y^2+\delta^2
      x^2}{1-\delta^2}}\sim \frac{y}{\tau_D }+\mathcal{O}(\delta^2)\,, \nonumber\\
  &&\sin \Phi_{12} = \frac{\vert \Gamma_{12}\vert^2 + 4 \vert
    M_{12}\vert^2 - (x^2+y^2)\vert q/p\vert^2/\tau_D^2}{4 \vert M_{12}
    \Gamma_{12}\vert}\sim \frac{x^2+y^2}{xy}\delta+\mathcal{O}(\delta^2)\,,
  \label{eq:m12g12}
\end{eqnarray}
with $\Phi_{12}=\arg(\Gamma_{12}/M_{12})$ and $\tau_D=0.41$
ps.
Consistently with the assumptions above, $\Gamma_{12}$ can be
taken real with negligible NP contributions, and a nonvanishing
$\Phi_{12}=-\Phi_{M_{12}}$ can be interpreted as a signal of new
sources of CP violation in $M_{12}$.

\begin{table}[t]
  \centering
  \begin{tabular}{|ccc|}
    \hline
    parameter & result @ $68\%$ prob. & $95\%$ prob. range\\
    \hline
    $\vert M_{12}\vert$ [ps$^{-1}$] & $(4.4 \pm 2.0) \cdot 10^{-3}$ & $[0.3,7.7]
    \cdot 10^{-3}$\\
    $\vert \Gamma_{12}\vert$ [ps$^{-1}$] & $(14.9 \pm 1.6) \cdot 10^{-3}$ &
    $[11.7,18.5] \cdot 10^{-3}$\\
    $\Phi_{M_{12}}$ [$^\circ$] & $(2.0 \pm 2.7)$ &
    $[-4,12]$\\
    \hline
    $\delta_{K\pi}$ [$^\circ$] & $(8 \pm 13)$  & $[-22, 30]$
    \\
    $\delta_{K\pi\pi}$ [$^\circ$] & $(-6 \pm 23)$  &
    $[-50,43]$ \\ \hline
    $x$ & $(3.6\pm 1.6) \cdot 10^{-3}$ & $[0.3,6.7]\cdot 10^{-3}$ \\
    $y$ & $(6.1\pm 0.7) \cdot 10^{-3}$ & $[4.8,7.6]\cdot 10^{-3}$ \\
    $\vert q/p\vert$ & $1.016\pm 0.018$ & $[0.981,1.058]$ \\
    $\delta$ & $(-1.6 \pm 1.8)\cdot 10^{-2}$ & $[-5.7,1.9] \cdot 10^{-2}$ \\
    $\phi [^\circ]$ & $-0.5\pm 0.6$ & $[-1.8,0.6]$\\
    $R_D$ & $(3.50\pm 0.04) \cdot 10^{-3}$ & $[3.43, 3.57]\cdot 10^{-3}$ \\
    $A_\Gamma$ & $(1.4\pm 1.5) \cdot 10^{-4}$ & $[-1.5, 4.4]\cdot 10^{-4}$ \\
    $R_M$ & $(2.4\pm 0.6) \cdot 10^{-5}$ & $[1.6,4.1]\cdot 10^{-5}$ \\
    $A_M$ & $(3.2\pm 3.6) \cdot 10^{-2}$ & $[-3.8, 11.3]\cdot 10^{-2}$ \\
    $y_\mathrm{CP}$ & $(6.1\pm 0.7) \cdot 10^{-3}$ & $[4.8, 7.6]\cdot 10^{-3}$ \\
  \hline
  \end{tabular}
  \caption{Results of the fit to $D$ mixing data.}
  \label{tab:ddmix_res}
\end{table}

\begin{figure}[htb]
  \centering
  \includegraphics[width=.3\textwidth]{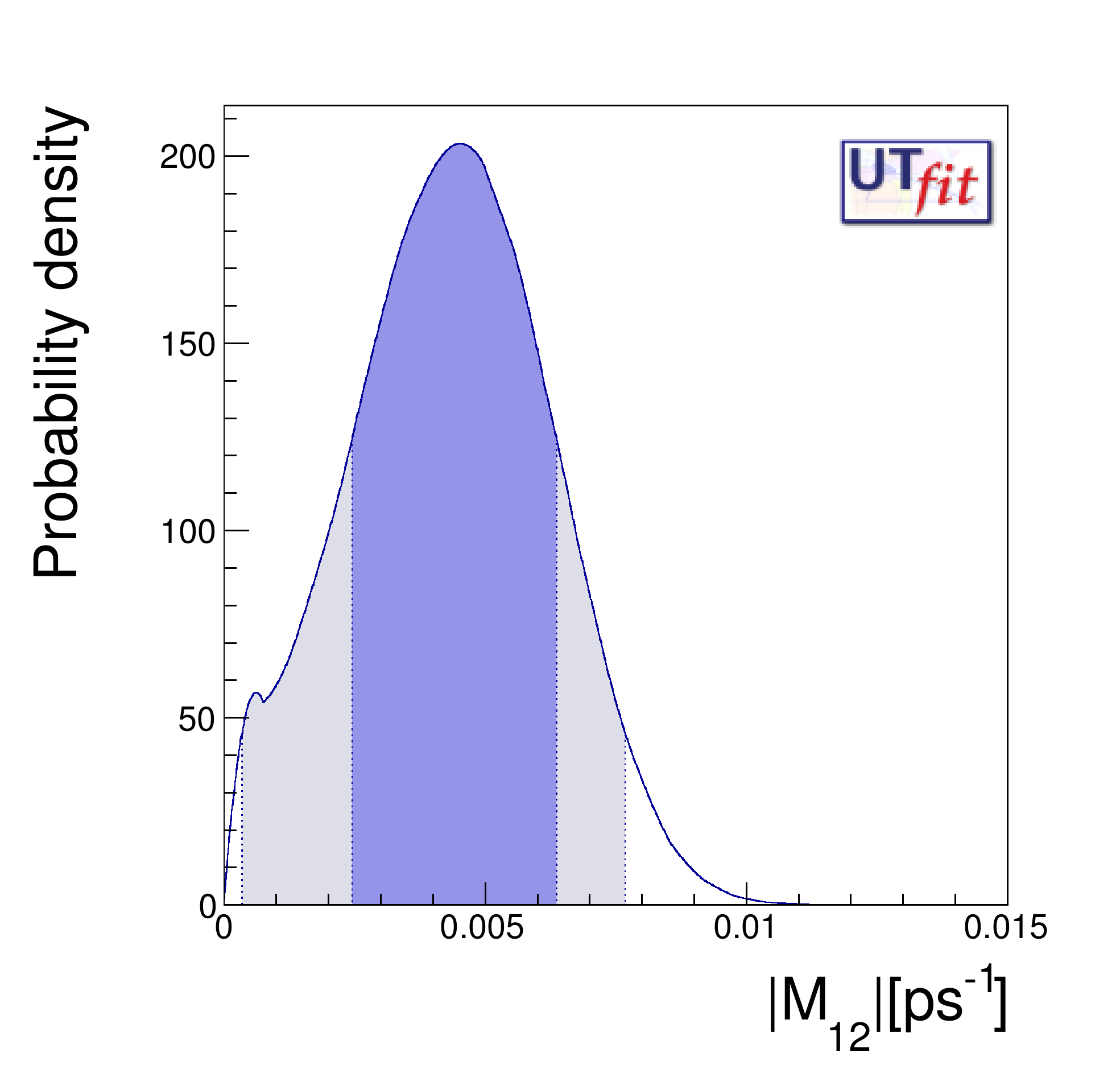}
  \includegraphics[width=.3\textwidth]{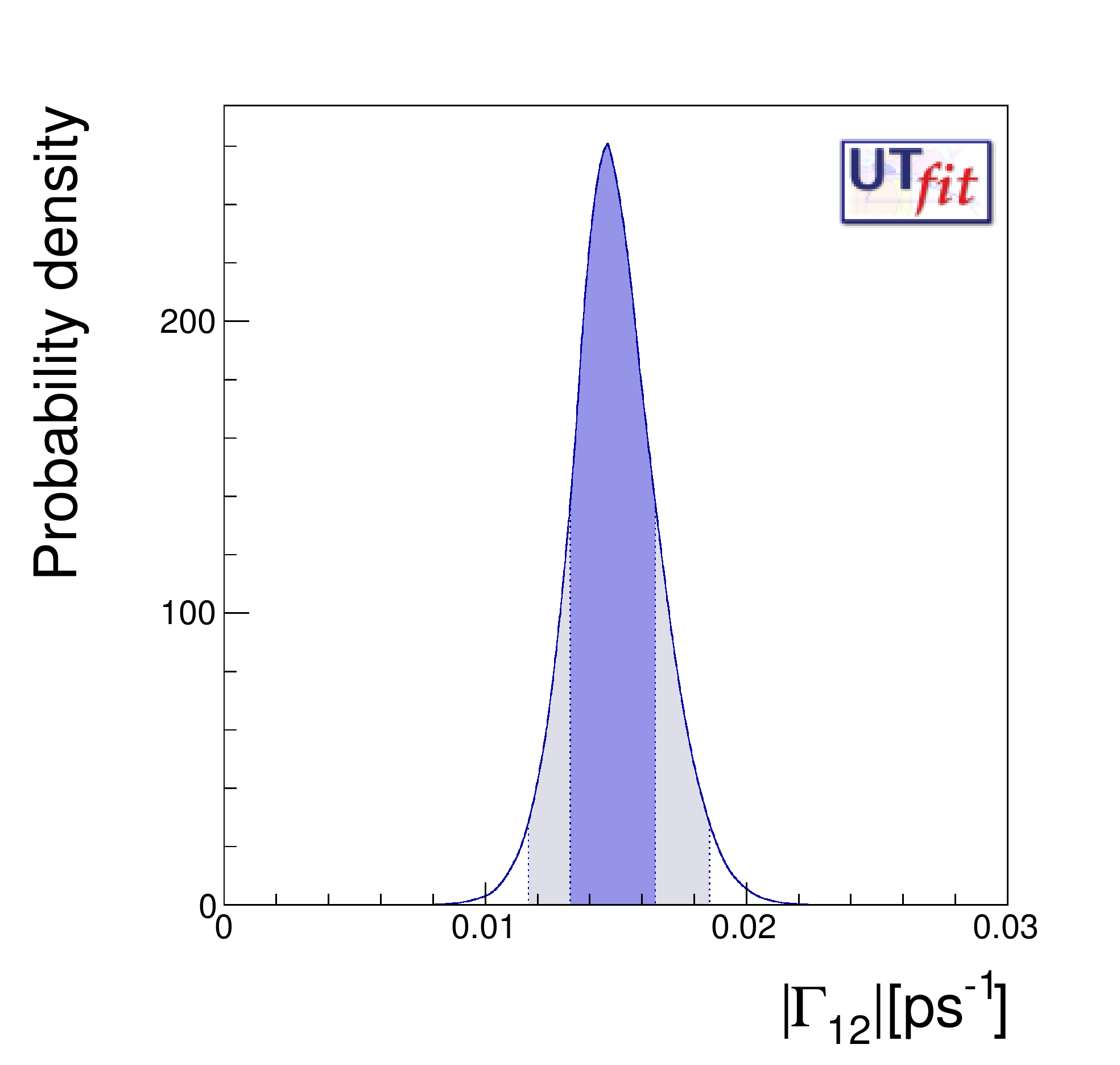}
  \includegraphics[width=.3\textwidth]{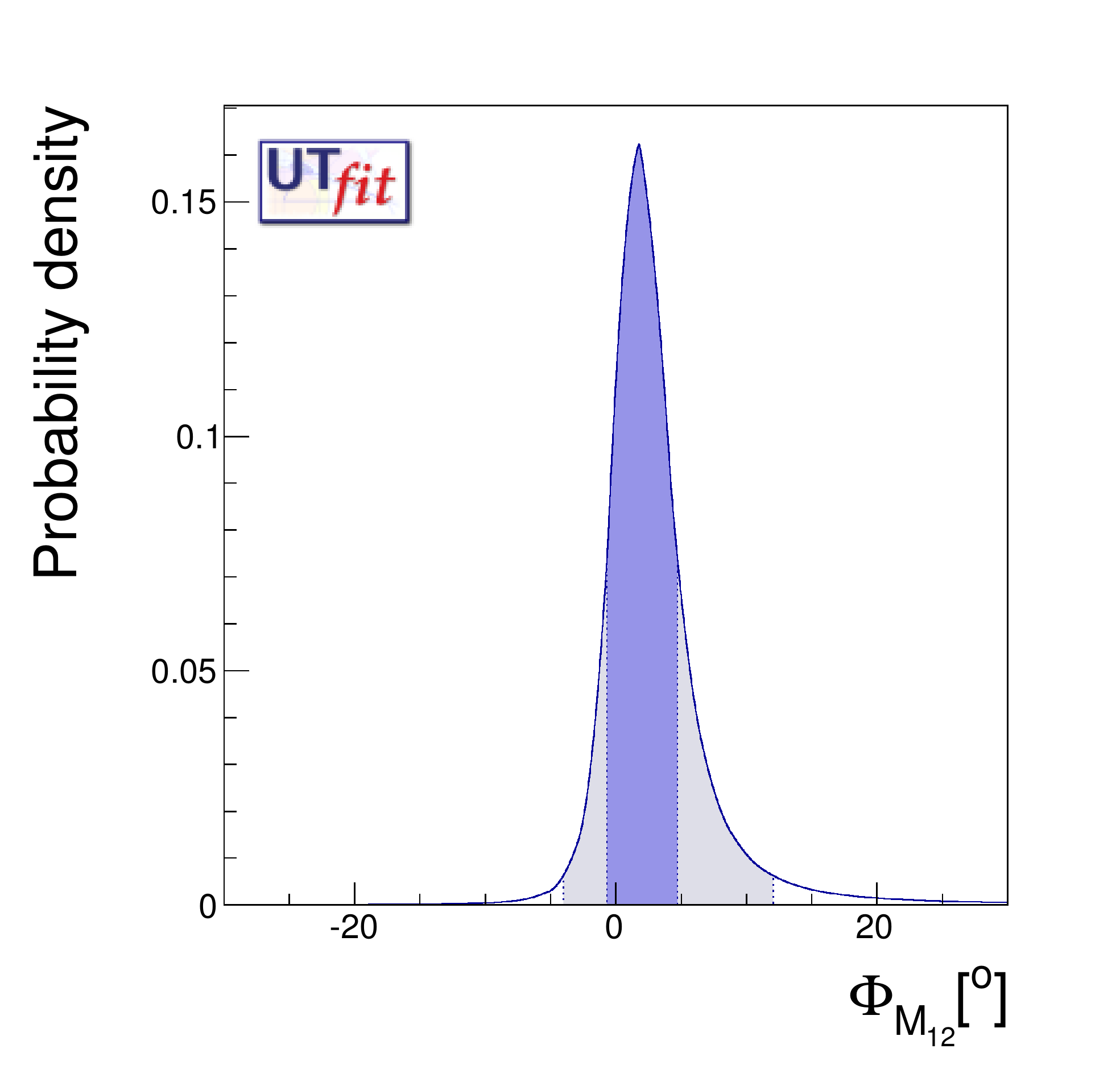}
  \caption{One-dimensional p.d.f. for the parameters $\vert M_{12}
    \vert$, $\vert \Gamma_{12}
    \vert$ and $\Phi_{M_{12}}$.}
  \label{fig:ddmix_1d}
\end{figure}

\begin{figure}[htb]
  \centering
  \includegraphics[width=.24\textwidth]{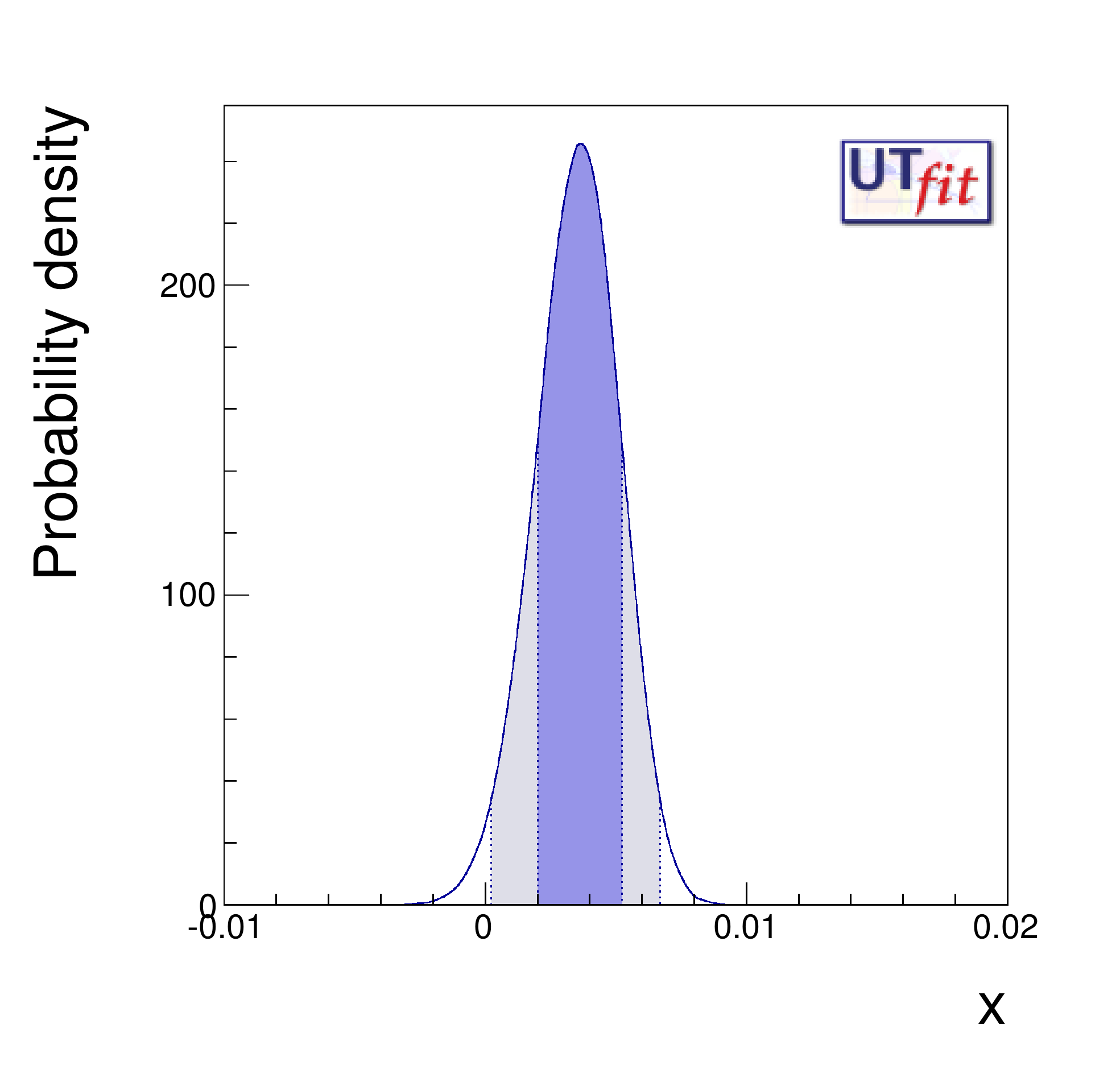}
  \includegraphics[width=.24\textwidth]{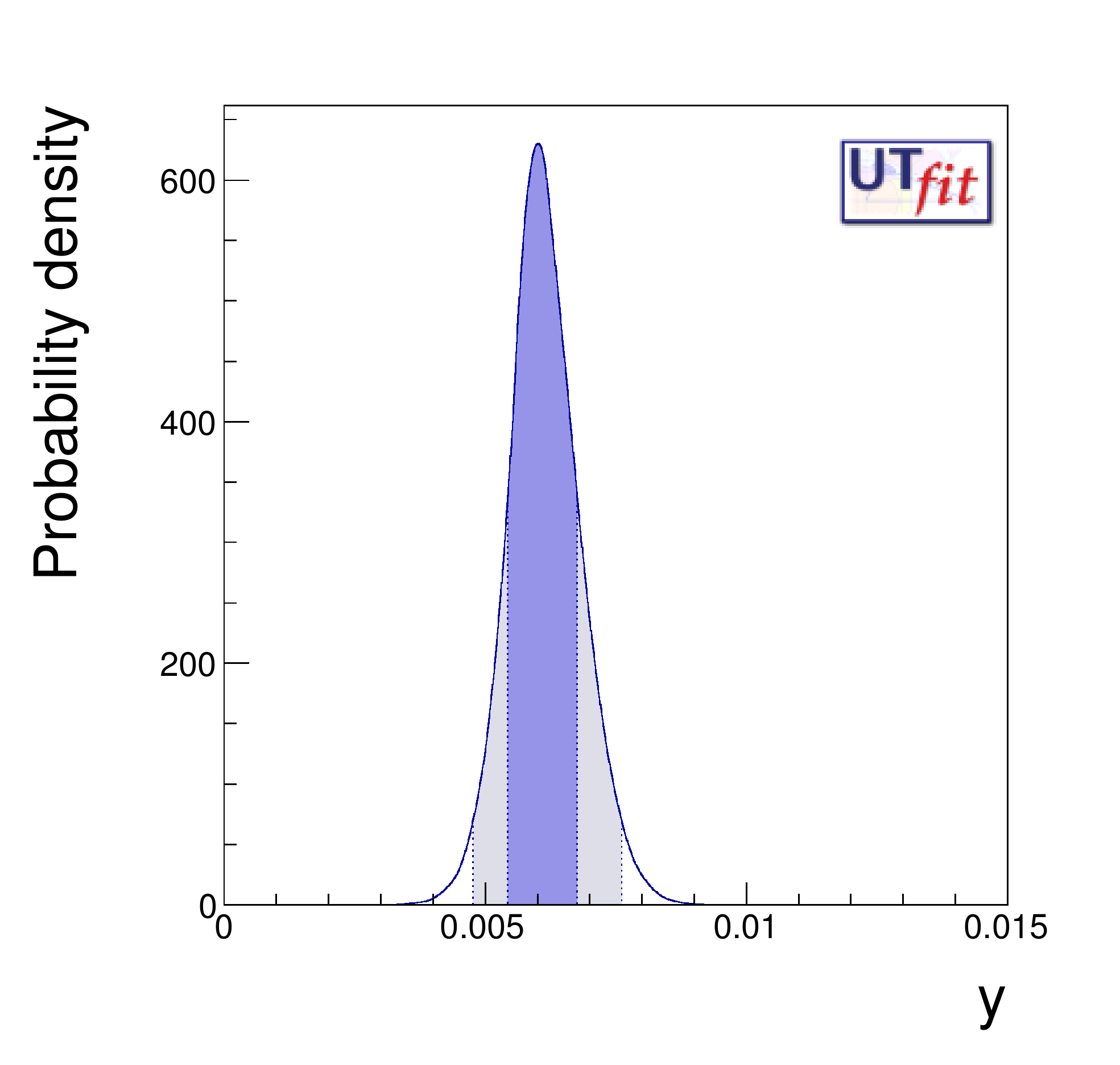}
  \includegraphics[width=.24\textwidth]{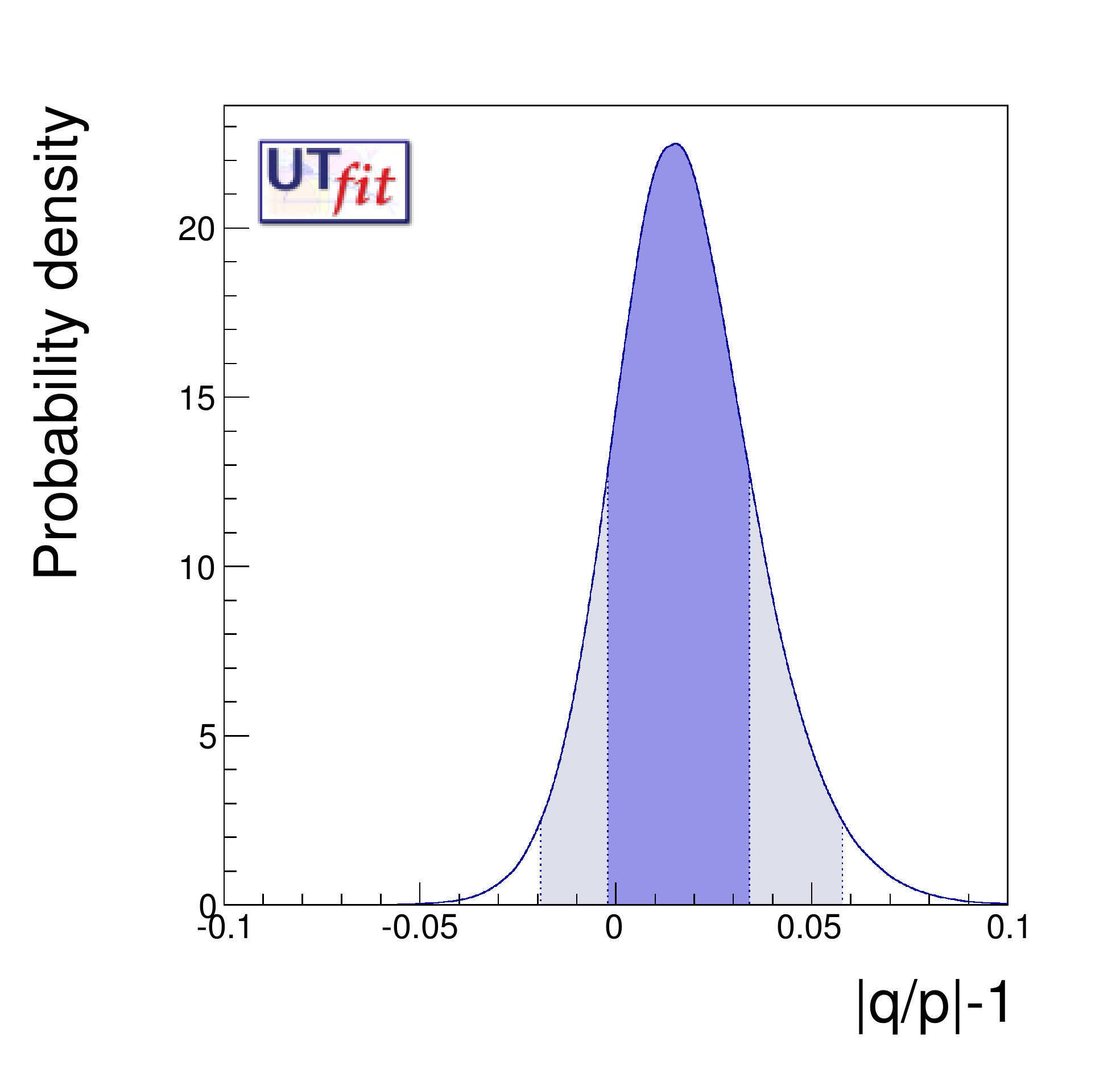}
  \includegraphics[width=.24\textwidth]{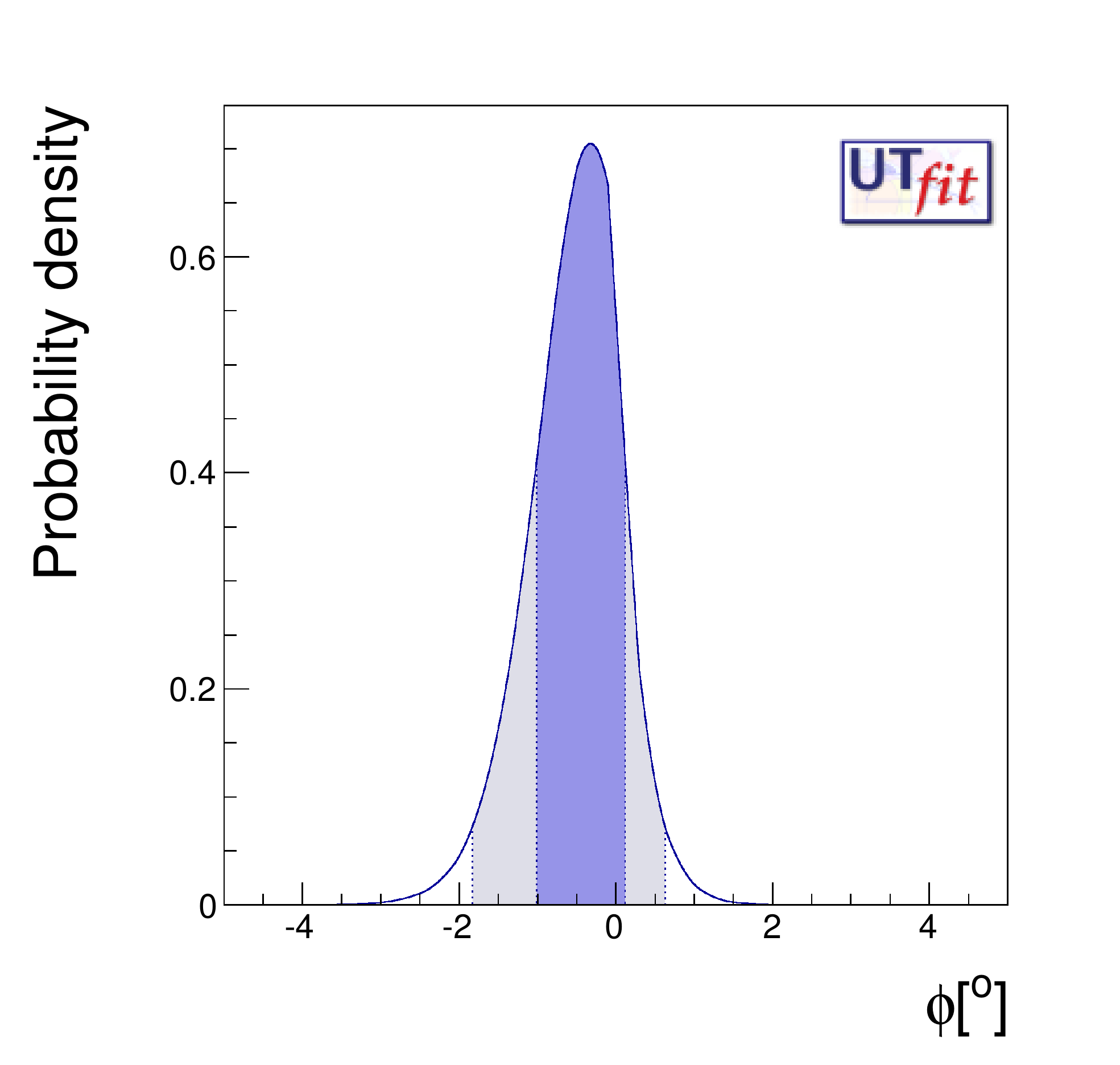}
  \caption{One-dimensional p.d.f. for the parameters $x$, $y$, $\vert
    q/p \vert -1$ and $\phi$.}
  \label{fig:ddmix_1d_2}
\end{figure}

\begin{figure}[htb]
  \centering
  \includegraphics[width=.3\textwidth]{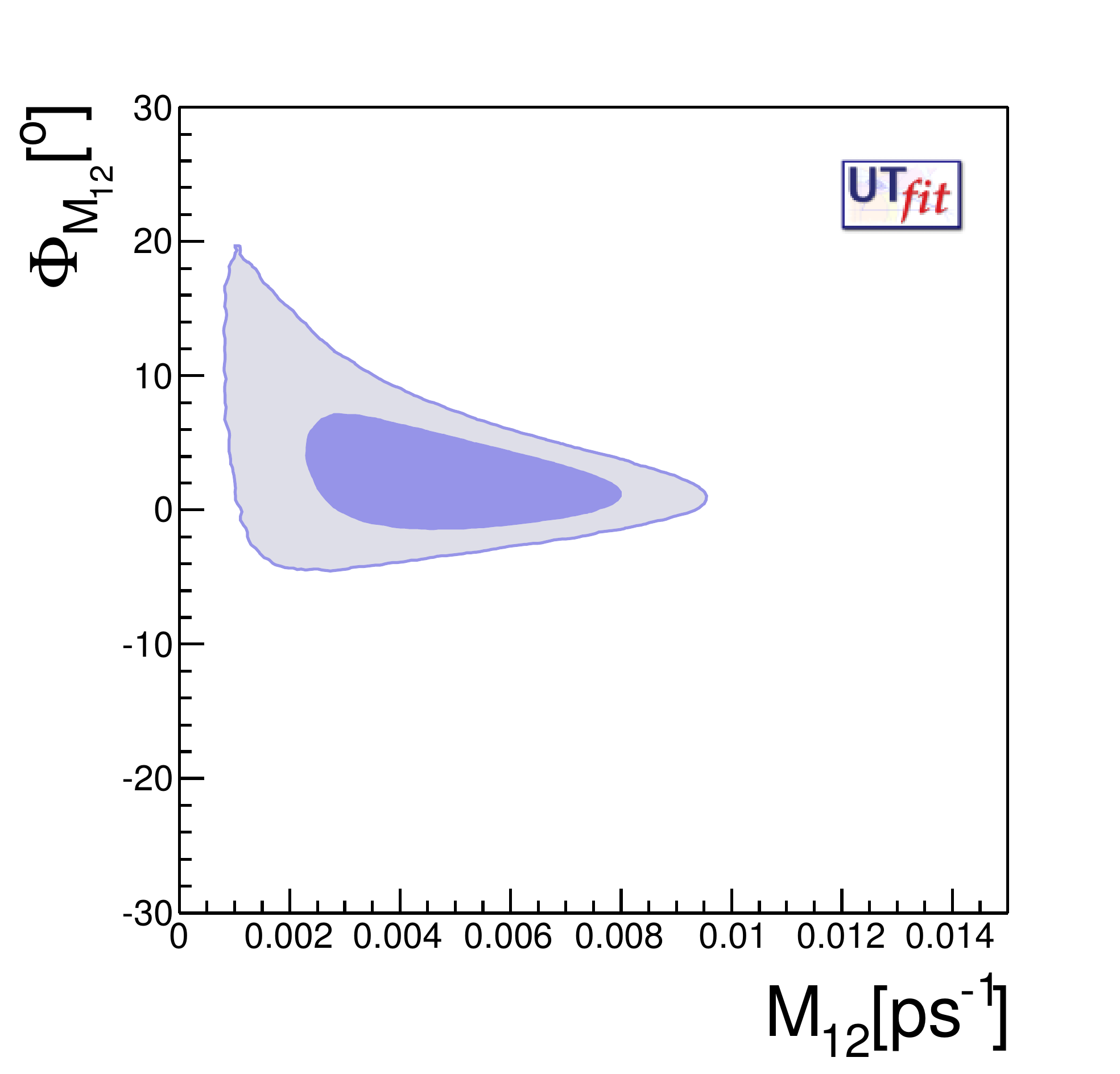}
  \includegraphics[width=.3\textwidth]{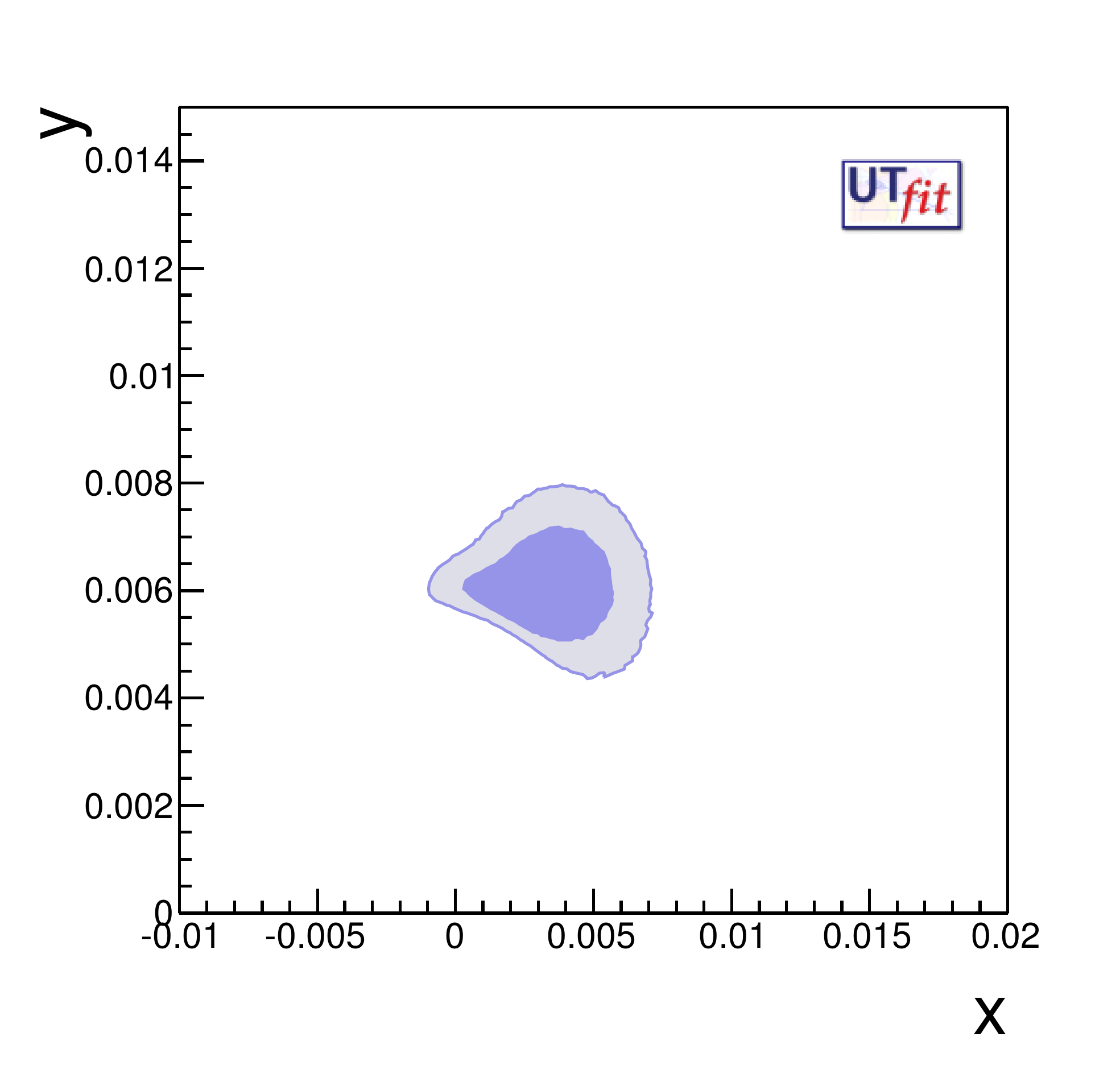}
  \includegraphics[width=.3\textwidth]{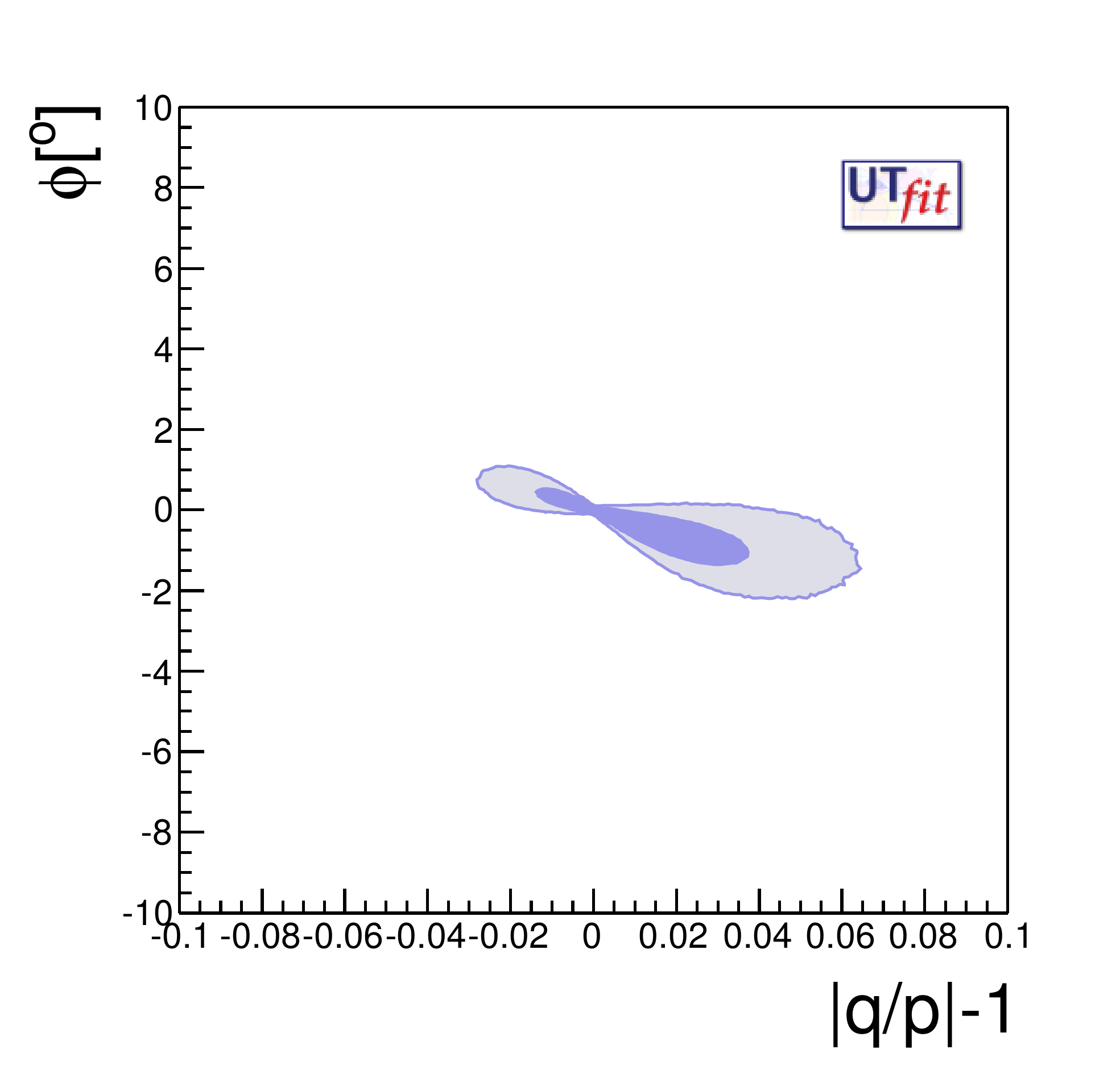}
  
  \caption{Two-dimensional p.d.f. for $\Phi_{12}$ vs
    $\vert M_{12} \vert$ (left), $y$ vs $x$ (middle) and
    $\phi$ vs $\vert q/p \vert -1$ (right).}
  \label{fig:ddmix_2d}
\end{figure}

The results of the fit are reported in Table \ref{tab:ddmix_res}. The
corresponding probability density functions (p.d.f.'s) are shown in
Figs. \ref{fig:ddmix_1d} and \ref{fig:ddmix_1d_2}. Some
two-dimensional p.d.f.'s are displayed in Fig. \ref{fig:ddmix_2d}.

As can be seen from Table~\ref{tab:ddmix_res}, the fitted value of
$\delta$ is at the percent level and indeed the central values of
$\vert M_{12}\vert$, $\vert \Gamma_{12}\vert$ and $\Phi_{12}$ are
compatible with the expanded formulae in eq.~(\ref{eq:m12g12}).
However in our fit we used the exact formulae since the region
of $x \lesssim 10^{-4}$, still allowed by experimental data (although with
probability less than 5\%), breaks the validity of the small $\delta$ expansion.

The results in Table \ref{tab:ddmix_res} can be used to constrain NP
contributions to $D - \bar D$ mixing and decays.

Our results are in very good agreement with the fit labeled ``No direct CPV
in DCS decays'' by HFAG \cite{[{}][{ and online
    updates at \url{http://www.slac.stanford.edu/xorg/hfag/}}]Amhis:2012bh},
now that HFAG uses the theoretical relation in eq.~(\ref{eq:phi}) as
we suggested in our previous paper.

\acknowledgments
M.C. is associated to the Dipartimento di Matematica e Fisica, Universit\`a di Roma
Tre. E.F. and L.S. are associated to the Dipartimento di Fisica,
Universit\`a di Roma ``La Sapienza''. The research leading to these results
has received funding from the European Research Council under the European 
Union’s Seventh Framework Programme (FP/2007-2013) / ERC Grant
Agreements n. 279972 ``NPFlavour'', n. 267985 ``DaMeSyFla'' and
from the People Programme (Marie Curie Actions) under European Union's
Seventh Framework Programme (FP7/2007-2013) / REA Grant
Agreement n. 329017 ``Charm@LHCb''.

\bibliography{hepbiblio}

\end{document}